\documentclass[twocolumn,showpacs,aps]{revtex4}

\usepackage{psfig}

\begin{document}

\title{Tensor to scalar ratio of perturbation amplitudes and
inflaton dynamics}

\author{C\'esar A. Terrero--Escalante}
\email{cterrero@fis.cinvestav.mx}
\affiliation{Instituto~de~F\'{\i}sica,~UNAM,
~Apdo.~Postal~20--364,~01000,~M\'exico~D.F.,~M\'exico.}%

\date{\today}

\begin{abstract}
For the inflaton
perturbations it is shown that the evolution 
of the difference between the
spectral indices can be translated into information
on the scale dependence of the tensor to scalar amplitudes ratio, 
$r$, and how the scalar field potential can be derived from
that information. Examples are given where $r$
converges to a constant value during inflation but dynamics
are rather different from the power--law model. 
Cases are presented
where a constant $r$ is not characteristic of the
inflationary dynamics though the resulting perturbation spectra are
consistent with the CMB and LSS data.

The inflaton potential corresponding to $r$ given by a n--th order polynomial
of the e--folds number is derived in quadratures expressions. 
Since the observable difference between the spectral
indices evaluated at a pivot scale yields information about the linear term 
of that polynomial, the first order case is explicitly written
down. The solutions show features beyond the exponential form corresponding to
power--law inflation and can be matched with current observational data. 
\end{abstract}

\maketitle

\section{\label{sec:Intro}Introduction\protect}
During the early phase of the cosmological evolution 
the universe could experience
a period of accelerated expansion known as inflation.
The simplest scenario with
a time dependent equation of state
yielding the required negative pressure 
is that of a single real scalar field, {\em the inflaton}, 
with dynamics dominated by its potential energy.
For extensive details and 
references on this scenario and on the other
topics mentioned in this Introduction see book \cite{inflation} and 
Ref.~\cite{Percival:2002gq}. 

The inflaton quantum fluctuations would be stretched by the expansion beyond
the radius within which causal interactions take place. 
These curvature perturbations could reenter the causal
horizon in much later epochs
and, through gravitational collapse, lead to 
anisotropies in the cosmic microwave background (CMB) temperature and
to cosmological large scale structure (LSS).
During inflation primordial gravitational waves are also produced.
These tensor perturbations induce 
a curled polarization in the CMB radiation and increase the overall amplitude
of its anisotropies at large angular scales.

A hint about the physics in the very early universe could be obtained
by fitting to data
the result of analytical calculations of the CMB and density 
spectra.
These calculations depend on the values of some parameters which are
the ones to be pinned down. The initial conditions
for the evolution of the thermal anisotropies 
are also characterized by several
quantities. These primordial parameters are often given as the multipoles of
\begin{eqnarray}
\label{eq:SExp}
\ln A^2(k)&=&\ln A^2(k_*) + n(k_*)\ln\frac{k}{k_*}
\nonumber\\
&+& \frac12\frac{d\,n(k)}{d\ln k}\vert_{k=k_*}\ln^2\frac{k}{k_*}+\cdots
\, ,
\end{eqnarray}
where $A$ stands for the normalized amplitudes of the scalar ($A_S$)
or tensor ($A_T$)
perturbations, the corresponding spectral indices, $n$, 
are defined by,
\begin{eqnarray}
\label{eq:nSDef}
n_S-1&\equiv&\frac{d\ln A_S^2}{d\ln k} \, ,\\
\label{eq:nTDef}
n_T&\equiv&\frac{d\ln A_T^2}{d\ln k} \, ,
\end{eqnarray}
$k=aH$ is the comoving wavenumber corresponding
to the wavelength matching
the Hubble horizon during inflation and $k_*$ is a pivot scale.

Initially, only $A_S(k_*)$ was fitted.
Since inflation predicts nearly scale--invariant
spectra, the tilt given by the scalar spectral index was then taken into 
account. This reduces the primordial spectrum to a power--law function 
of the scale $k$. The only single field model exactly yielding such a spectrum
is power--law inflation \cite{PLinfl} where the cosmic scale factor, $a$, 
behaves like a
power--law of the cosmic time, $t$, and the inflaton potential is an 
exponential function.
Signals of nonzero `running'  of the scalar index, 
$dn_S /d\ln k$, have already been reported \cite{Hannestad:2001nu}
and must be refined by near future observations, 
allowing, this way, to move beyond the power--law approximation.

The role of the tensor perturbations
deserves also attention when determining the 
best--fit values of the cosmological parameters 
from CMB and LSS spectra. 
That is motivated in part by the possibility of 
measuring the cosmic background polarization,
allowing the tensorial contribution 
to be indirectly determined. 
This contribution can be parametrized in terms of 
the relative amplitudes 
of the tensor and scalar perturbations,
\begin{equation} 
\label{eq:r}
r \equiv \alpha \frac{A_T^2}{A_S^2}\, ,
\end{equation}  
where $\alpha$ is a constant. 
Presently,
due to measurement limitations, a constant value of $r$ is fitted.

Nevertheless, in Ref.~\cite{Terrero-Escalante:2001du} it was shown that 
few inflationary models 
produce an exactly constant tensor to scalar ratio, 
and, in order to be proper scenarios of inflation, they must be
observationally indistinguishable from power--law inflation, 
where $r={\rm constant}$ too. 
Since power--law inflation is just one of many suitable final
stages of the inflaton dynamics \cite{Terrero-Escalante:2002sd}, 
it
implies that, 
either a constant value for $r$ may be no characteristic of the
underlying inflationary evolution
or that
conclusions about the inflaton potential beyond an exponential form may 
be no possible to be drawn \cite{Terrero-Escalante:2002qe}. 
This last factor could
be a strong limitation for programs of the inflaton potential reconstruction.

Taking the above into account, the aim of this letter is to analyze
the relation between
the functional form of the tensor to scalar ratio and the inflationary 
dynamics. After introducing in Sec.~\ref{sec:keq} 
the relevant equations, the analysis
is done in Sec.~\ref{sec:sols} by means of several examples.
Finally, the results are discussed in Sec.~\ref{sec:disc} and
it is concluded that the difference between the spectral indices can be
a very useful quantity because it yields information on the scale dependence
of the tensor to scalar ratio, hence allowing to observe features of the
inflaton potential different from the exponential form.

\section{\label{sec:keq}Key equations}

The horizon--flow functions proved to be useful quantities while
describing the inflationary dynamics and perturbations. They were defined 
in Ref.~\cite{HFFampl} as
\begin{equation}
\label{eq:Hjf}
\epsilon_0 \equiv \frac{{d_{\rm H}}(N)}{d_{\rm Hi}}, \qquad
\epsilon_{m+1} \equiv {{\rm d} \ln |\epsilon_m|
\over {\rm d} N}, \qquad m\geq 0 \, ,
\end{equation} 
where $d_{\rm H} \equiv a/\dot{a}$ denotes the Hubble distance, 
$N \equiv \ln (a/a_{\rm i})$ the number of e--folds since time 
$t_{\rm i}$, and $d_{\rm Hi} \equiv d_{\rm H}(t_{\rm i})$. During inflation,
$0\leq\epsilon_1<1$. For $m>1$, $\epsilon_m$ may take any real value, though
usually
$|\epsilon_m|<1$ for the inflationary spectra to be weakly scale dependent.
For instance, in Ref.~\cite{Leach:2002dw} the CMB constrains 
$\epsilon_1<0.1$ and $|\epsilon_2|<0.3$ were found for fixed $\epsilon_3=0$.

To next--to--leading order in an expansion in
terms of the horizon--flow functions the spectral indices 
(\ref{eq:nSDef}) and (\ref{eq:nTDef}) 
can be written as
\cite{Stewart:1993bc,Terrero-Escalante:2002sd}
\begin{eqnarray}
\label{eq:SLns}
n_S-1&=& -2\epsilon_1 - \epsilon_2 - 2\epsilon_1^2 - (2C+3)\epsilon_1\epsilon_2
- C\epsilon_2\epsilon_3, \\
\label{eq:SLnt}
n_T&=& -2\epsilon_1 - 2\epsilon_1^2 - 2(C+1)\epsilon_1\epsilon_2 \, ,
\end{eqnarray}
where $C\approx-0.7293$.
If these expressions are analyzed as a system of differential
equations (where the differentiation, according
with definitions (\ref{eq:Hjf}), is done with respect to $N$), 
then it is not a closed system. 
Even after differentiating Eq.~(\ref{eq:SLnt})
and substituting the result and 
Eq.~(\ref{eq:SLnt}) itself into 
Eq.~(\ref{eq:SLns}), one 
still has to deal with three independent variables, namely, $\epsilon_1$,
$n_S$ and $n_T$ 
(see Refs.~\cite{Ayon-Beato:2000,Terrero-Escalante:2002sd} 
for more detailed discussion). There 
are not other
independent equations relating these variables which could be used to close the
system of equations (\ref{eq:SLns}) and (\ref{eq:SLnt}). Therefore, 
information on the functional forms of the observables $n_S$ and 
$n_T$ it is necessary in order to describe the dynamics of $\epsilon_1$. This
is the basic philosophy behind the Stewart--Lyth inverse problem 
\cite{Ayon-Beato:2000}.

It follows from definitions 
(\ref{eq:nSDef}), (\ref{eq:nTDef}) and (\ref{eq:r})
that
\begin{equation}
\label{eq:dlnr/dlnk}
\frac{d\ln r}{d\ln k}= \Delta n \equiv n_T - (n_S-1) \, .
\end{equation}
This way, any information on the evolution of both spectral indices can be
used as information on the scale dependence of the tensor to scalar ratio.
Using expression (\ref{eq:dlnr/dlnk}), after 
subtracting Eq.~(\ref{eq:SLns}) to Eq.~(\ref{eq:SLnt}) yields
\begin{equation}
\label{eq:dlnrdlnk1}
\frac{d\ln r}{d\ln k}= \epsilon_2 + \epsilon_1\epsilon_2 +
C\epsilon_2\epsilon_3 \, .
\end{equation}
Equality $k=aH$ implies,
\begin{equation}
\frac{dN}{d\ln k}=\frac{1}{1-\epsilon_1}
\, .
\label{eq:g}
\end{equation}
Then, changing variables in Eq.~(\ref{eq:dlnrdlnk1}) it is obtained,
\begin{equation}
\label{eq:dlnrdN}
C\frac{d\epsilon_2}{dN} + \epsilon_2 = \frac{d\ln r}{dN} \, ,
\end{equation}
where Eqs.~(\ref{eq:Hjf}) were used and 
terms with 
order higher than the second 
one were neglected 
consistently with the approximations
applied to derive Eqs.~(\ref{eq:SLns}) and (\ref{eq:SLnt}) 
\cite{Stewart:1993bc}.

Taking into account the definition (\ref{eq:Hjf}) for $\epsilon_2$, 
Eq.~(\ref{eq:dlnrdN}) can be rewritten as,
\begin{equation}
\label{eq:lnr}
C\frac{d\ln \epsilon_1}{dN} + \ln \epsilon_1 = \ln\frac r {r_0} \, ,
\end{equation}
where $r_0$ is an integration constant. 
Now, the number of variables to solve for has been reduced to two by
comprising in $r$ the information on $n_S$ and $n_T$. Remains only
one equation to deal with thus, it is still 
compulsory to find out how
$r$ behaves to describe the dynamics of $\epsilon_1$.  

According with definitions (\ref{eq:Hjf}),
\begin{equation}
\label{eq:H^2}
H^2(N) = H_0^2\exp\left(-2\int \epsilon_1(N) dN \right)
\, ,
\end{equation}
where $H_0$ is an integration constant. 
On the other hand using the definition (\ref{eq:Hjf}) for $\epsilon_1$ and
the Friedmann equation for the inflaton cosmology
\cite{inflation,Terrero-Escalante:2001rt},
\begin{equation}
\frac{d\phi}{dN}=\sqrt{\frac 2 \kappa}\sqrt{\epsilon_1} \, ,
\end{equation}
where $\phi$ is the inflaton,
$\kappa\equiv 8\pi/m^2_{\rm Pl}$ and $m_{\rm Pl}$ is the Planck mass.
Hence, given a solution for $\epsilon_1(N)$ the
corresponding potential as function of the inflaton field is,
\begin{equation}
\label{eq:Vphi}
V(\phi)= \left\{
\begin{array}{rcl}
\phi(N)&=& \sqrt{\frac2\kappa}{\displaystyle \int} 
\sqrt{\epsilon_1} dN - \phi_0
\, , 
\\
V(N)&=&\frac{H_0^2}\kappa\left[3-\epsilon_1\right]
\exp\left(-2{\displaystyle \int} \epsilon_1 dN \right)
\, ,
\end{array}
\right.
\label{eq:C2Vp}
\end{equation} 
where the expression for the potential as function of $N$
is derived from the Einstein
equations for the scalar field cosmology \cite{inflation,Ayon-Beato:2000}.
In this way, the inflaton potential can, in principle, 
be determined from just the information
on the functional form of the tensor to scalar ratio or, equivalently,
on the evolution of the difference between the tensor and scalar
spectral indices.

\section{Some solutions}
\label{sec:sols}

Current analysis of CMB and LSS observations 
yields $r<1$ 
\cite{inflation,Leach:2002dw}, 
hence solutions satisfying $\ln r<0$ shall be searched for. 
Eq.~(\ref{eq:lnr}) can be written as,
\begin{equation}
\label{eq:re1e2}
r = r_0\epsilon_1\exp(C\epsilon_2) \, .
\end{equation}
Expanding this expression to next--to--leading order and comparing
with the corresponding expansion for $A_T/A_S$ \cite{Stewart:1993bc}
it can be determined that $r_0=16$ \cite{HFFampl,Leach:2002dw}. 
For large values of $|\epsilon_2|$, models with $\epsilon_2>0$ are favored. 
For the next--to--leading order approximation to proceed it must be required
the minimum value, $\epsilon_2=-0.3$. Together with the constrain $r<1$,
this lead to $\epsilon_1<0.05$. These values coincides with the 
$2\sigma$ bounds
reported in Ref.~\cite{Leach:2002dw}.

Both,
Eqs.~(\ref{eq:dlnrdN}) and (\ref{eq:lnr}) are in the class of 
first order linear differential equations like,
\begin{equation}
\label{eq:tempEq}
y\prime + \alpha y = g(x) \, ,
\end{equation}
with a prime denoting differentiation with respect to $x$ and $\alpha$ a 
constant.
Using the ansatz
$y=\exp(-\alpha x)f(x)$, the general solution of Eq.~(\ref{eq:tempEq}) is 
obtained as,
\begin{equation}
\label{eq:tempSol}
y = \exp(-\alpha x)\left[B + \int_{x_0}^x g(x)\exp(\alpha x) dx \right]\, ,
\end{equation}
where $B$ is the constant arising from the integration of the homogeneous 
equation and the integration limits are determined by the range
of $x$ where $g(x)$ is continuous. 
Thus, one has the option of solving any of Eqs.~(\ref{eq:dlnrdN}) or
(\ref{eq:lnr}), regarding the complexity
of the corresponding left hand side expression. 
If $g(x)\rightarrow \beta= {\rm const.}$ 
while $x\rightarrow \infty$, then, for $\alpha>0$, 
applying the L'Hopital's rule it can 
be determined that, for any lower integration limit, the solution converges to
$\beta/\alpha$. However, for $\alpha<0$ the only solution converging to 
$\beta/\alpha$ is,
\begin{equation}
\label{eq:rareSol}
y = -\exp(-\alpha x)\int_x^{\infty} g(x)\exp(\alpha x) dx \, .
\end{equation} 
Since in Eqs.~(\ref{eq:dlnrdN}) and (\ref{eq:lnr}), 
$\alpha=1/C<0$, this implies an attractor given by 
the particular solution $\epsilon_1=r_*/16$ 
for $r = r_* = {\rm const.}$ (or
$\Delta n = 0$ from Eq.~(\ref{eq:dlnr/dlnk})). This 
corresponds to power--law inflation
and it is just a special case
(that for $B=0$), 
even if $r\rightarrow r_*$ while
$N\rightarrow \infty$. For instance,
if it is assumed 
\begin{equation}
\label{eq:Sol1}
\ln r = b \exp(-a N) + \ln r_\infty \, ,
\end{equation} 
with $r_\infty$, $a>0$ and $b$ some constants then, 
according with
Eq.~(\ref{eq:rareSol}), the asymptotic solution for $\epsilon_1$ 
will be
\begin{equation}
\label{eq:pl1pc}
\epsilon_1 = \epsilon_{1\infty} \exp\left[
-\frac{b}{C a-1}\exp(-a N)
\right] \, ,
\end{equation}
where $\epsilon_{1\infty}=r_\infty/16$. This solution converges to 
$\epsilon_1=\epsilon_{1\infty}$ while $N\rightarrow \infty$. 
For $b>0$ ($b<0$), 
this happen starting from values larger (smaller) than $\epsilon_{1\infty}$. 
If $b=0$, 
corresponding to $r=r_\infty={\rm const.}$, 
then the outcome is just the power--law
solution given by $\epsilon_1(N)=\epsilon_{1\infty}$.
This is the kind
of behavior observed in Ref.~\cite{ConstNt} 
for models with constant tensorial index
and also
in the first case analyzed in 
Ref.~\cite{Terrero-Escalante:2001rt}
for models with weakly scale dependent spectral indices. 

According to Eq.~(\ref{eq:rareSol}), the more general solution 
with $r$ given by
Eq.~(\ref{eq:Sol1}) is,
\begin{equation}
\label{eq:pl1gc}
\epsilon_1 = \epsilon_{1\infty} 
\exp\!\!\left[
B\exp(-\frac1C N)
\right]
\exp\!\!\left[
-\frac{b}{C a-1}\exp(-a N)
\right].
\end{equation} 
Now the dynamics is different. 
If $B>0$ ($B<0$) the solution increases 
(decreases) very fast starting from,  
\begin{equation}
\label{eq:pl1gce10}
\epsilon_1 = \epsilon_{1\infty} 
\exp\left(B\right)
\exp\left(-\frac{b}{C a-1}\right) \, .
\end{equation}
If $b=0$, this behavior is modified 
only in the rate of change of $\epsilon_1$,
consistently with the solution in 
Ref.~\cite{Terrero-Escalante:2001du} for models with $r={\rm const.}$.
Thus, $B$ is a bifurcation parameter for the inflationary dynamics
described by Eq.~(\ref{eq:pl1gc}).

Functional forms of $r$ converging to a constant value are motivated 
by difficulties of measuring the primordial gravitational waves background
\cite{inflation}.
Maybe the unique for a while
alternative for determining the contribution
of tensor modes to CMB is the derivation of the tensor to scalar ratio 
from the background radiation polarization, 
the more optimistic 
expectations being measuring a constant central value of $r$
\cite{inflation}. 
However, there are other behaviors of $r$ consistent with
current constrains on the primordial spectra. For example, if in 
Eq.~(\ref{eq:Sol1}) $a$ is changed for $-a$, the general solution 
(\ref{eq:pl1gc}) is modified in exactly the same way, i.e., $a$ for $-a$.
Nevertheless, now we have richer dynamics as for $r$ as for $\epsilon_1$,
depending on the values of the different parameters involved. For $B=0$ and
$b<0$, $r\rightarrow 0$ with $N\rightarrow \infty$ but there are two different
behaviors for $\epsilon_1$; if $0<a<-1/C$, $\epsilon_1\rightarrow \infty$, 
else $\epsilon_1\rightarrow 0$. Interesting is the case $B=0$ and
$b>0$, since $r\rightarrow \infty$ with $N\rightarrow \infty$ but for
$0<a<-1/C$, $\epsilon_1\rightarrow 0$. According to Eq.~(\ref{eq:SLns}),
this example
can be made consistent with current analysis
of CMB and LSS observations 
\cite{Percival:2002gq,Hannestad:2001nu,Leach:2002dw}
by tunning the values 
of $a$ and $b$ to control the behavior of $\epsilon_2$ and $\epsilon_3$.
For $B\neq 0$ more possibilities arise depending on 
whether $B$ is positive or negative, and greater or less than $b/(1+aC)$.
Solutions with  $r\rightarrow \infty$, $\epsilon_1\rightarrow 0$ 
 and $|\epsilon_2|<1$, $|\epsilon_3|<1$ while 
$N\rightarrow \infty$, can be 
easily found.

Equally satisfactory, from the observational point of view, are functional
forms of $r$ exhibiting a transient period of almost power--law behavior 
during a sufficiently large numbers of $N$. For example, 
a proper choice of the constants $a$, $b$, $\alpha$, $\beta$ ensures that
to be the case for
\begin{eqnarray}
\label{eq:qsplr}
r &=& r_0\left\{a\tan[\alpha(N-N_0)]-b\right\}
\nonumber \\
&\times& \exp\left\{\frac{\beta}{cos^2[\alpha(N-N_0)]
\left(a\tan[\alpha(N-N_0)]-b\right)}\right\} \nonumber  
\end{eqnarray}
and the corresponding solution for $\epsilon_1$,
\begin{equation}
\label{eq:qsple1}
\epsilon_1 = \epsilon_{1(0)}\left\{a\tan[\alpha(N-N_0)]-b\right\}
\, .
\end{equation}
This result is similar to case 2 in 
Ref.~\cite{Terrero-Escalante:2001rt}
for models with weakly scale dependent spectral indices.

Now, if $g(x)$ in Eq.~(\ref{eq:tempEq}) is given 
precise enough by,
\begin{equation}
\label{eq:gx}
g(x)=\sum_{p=0}^n a_p(x-x_0)^p
\, ,
\end{equation}
with $n$ some finite integer,
then solution (\ref{eq:tempSol}) reads,
\begin{eqnarray}
\label{eq:gSol}
y(x)&=&B_0\exp(-\alpha x) \nonumber \\
&+& 
\sum_{p=0}^n\, \sum_{q=0}^p a_p\, (-1)^q \frac{p\,!}{(p-q)\,!}\,\alpha^{-(q+1)}
\, (x-x_0)^{p-q} \nonumber
\end{eqnarray}
for $p\geq q$.
The solution of Eq.~(\ref{eq:lnr}) is then,
\begin{eqnarray}
\label{eq:gSole1}
\epsilon_1(N)&=& \epsilon_{1(0)} \exp\left[B_0\exp(-\frac1C N)\right]
\nonumber \\
&\times&  
\prod_{p=0}^n\, \prod_{q=0}^p 
\exp\left[b_{p,q}\, (N-N_0)^{p-q}\right]
\, ,
\end{eqnarray}
where $b_{p,q}=a_p\, (-1)^q \,C^{(q+1)}{p\,!}/{(p-q)\,!}$.
The general expression for the inflaton potential
corresponding to $r$ given by a n--th order polynomial
of the e--folds number, $N$, can be derived in terms of 
quadratures expressions for $\epsilon_1$, by 
substituting solution (\ref{eq:gSole1}) into Eq.~(\ref{eq:Vphi}).
 
\begin{figure}[htb]
\centerline{\psfig{file=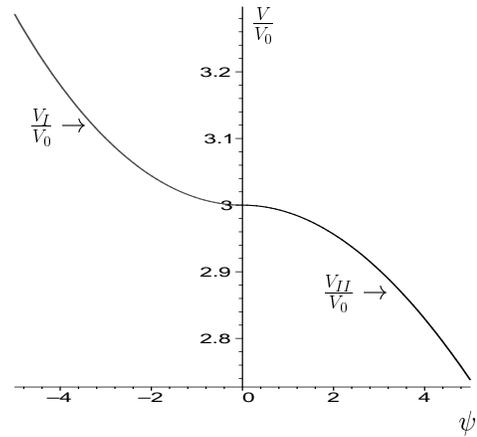,width=7cm}} 
\caption{Sectors of the inflaton potential given by Eq.~(\ref{eq:Vai}) for
$A=-0.0073$ ($V_I$) and $A=0.0073$ ($V_{II}$).}
\label{fig:01}
\end{figure}

\begin{figure}[htb]
\centerline{\psfig{file=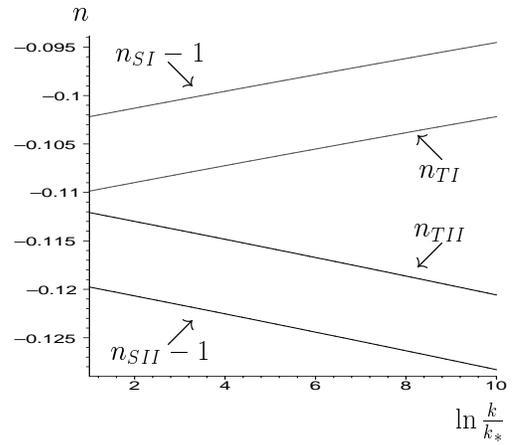,width=7cm}} 
\caption{The spectral indices for $V_I$ and $V_{II}$ with
$\epsilon_{1(i)}=0.05$.}
\label{fig:02}
\end{figure}

Unfortunately, the future prospects for obtaining information from 
observations for higher order coefficients in expansion 
(\ref{eq:gx}) are strongly limited. However, it is timely to recall that 
$dr/d\ln k = r\Delta n$ which, according to Eqs.~(\ref{eq:dlnrdlnk1}) and
(\ref{eq:re1e2}) is a next--to--leading
order expression, hence $a_1\approx a_0\Delta n(N_0)$.
With these regards, it becomes important to analyze in details case,
\begin{equation}
\label{eq:ra1}
\ln\frac{r}{16} = a_0 + a_1 (N-N_0)\, ,
\end{equation} 
where the corresponding solution for $\epsilon_1$ is,
\begin{equation}
\label{eq:e1a1}
\epsilon_1(N) = \epsilon_{1(i)}
\exp\left[B\exp\left(-\frac N C\right)\right]\exp\left(A N\right)
\, ,
\end{equation}
with $\epsilon_{1(i)}\equiv \epsilon_{1(0)}
\exp\{\left[a_0+\left(N_0-C\right)a_1\right]C\}$ and $A\equiv a_1 C$.
The asymptotes of this solution for $B\neq 0$ 
will be mainly determined by the value and sign of B. 
However, in the same fashion as it was shown in
Ref.~\cite{Terrero-Escalante:2001du} for $A=0$ (i.e., $\ln r/16=a_0$),
if the model yielding $\epsilon_1$ given by Eq.~(\ref{eq:e1a1})
is expected to be compatible with current data, 
$B$ has to be chosen an
extraordinarily small number, making the corresponding scenario 
very difficult to be observationally distinguished from
power--law inflation.
More interesting is the case $B=0$. 
From Eq.~(\ref{eq:Vphi}),
the corresponding inflaton potential is obtained as,
\begin{equation}
\label{eq:Vai}
V = V_0\left(3-\frac {A^2} 4 \psi^2\right)\exp\left(-\frac A 2 \psi^2\right)\,,
\end{equation}
where $\psi\equiv \sqrt{\kappa/2}(\phi+\phi_0)$. In Fig.~\ref{fig:01}
sectors of
this potential are plotted for $A=-0.0073$ ($V_I$) and $A=0.0073$ ($V_{II}$).
In Fig.~\ref{fig:02} 
are presented the corresponding to this model
scalar and tensorial spectral indices,
\begin{eqnarray}
\label{eq:nSai}
n_S &=& 1 - A + \left[2+(2C+3)A\right]
{\mathcal L}_W\left[-\epsilon_{1(i)}\left(\frac k {k_*}\right)^A\right]
\nonumber \\
&-& 2{\mathcal L}_W^2\left[-\epsilon_{1(i)}\left(\frac k {k_*}\right)^A\right]
,
\\
n_T &=&  2\left[1+(C+1)A\right]
{\mathcal L}_W\left[-\epsilon_{1(i)}\left(\frac k {k_*}\right)^A\right]
\nonumber \\
&-& 2{\mathcal L}_W^2\left[-\epsilon_{1(i)}\left(\frac k {k_*}\right)^A\right]
,
\end{eqnarray}
where Eqs.~(\ref{eq:SLns}), (\ref{eq:SLnt}) and (\ref{eq:g}) were used
and ${\mathcal L}_W[x]$ is the Lambert $W$ function \cite{LW}.
For a large set of $A$ and
$\epsilon_{1(i)}$ values,
these spectra agree with CMB and LSS observations, 
$n_S=1.033\pm 0.66$, $n_T=0.09\pm 0.16$ \cite{Percival:2002gq},
$-0.05<d n_S/d\ln k<0.02$ \cite{Hannestad:2001nu},
$\epsilon_1<0.05$ and $12.5\epsilon_1<r<1$ \cite{Leach:2002dw}.

\section{\label{sec:disc}Discussion}

The analysis of the general solution 
(\ref{eq:tempSol})
confirms
that the power--law inflationary scenario is just one among many
suitable final stages for the inflaton dynamics as it was previously
discussed in Ref.~\cite{Terrero-Escalante:2002sd}.
Moreover, as the examples in the
previous section indicate, a tensor to scalar ratio $r$ converging
to a constant value while the early universe inflates
is neither 
a necessary nor a sufficient condition
for the corresponding inflationary model to 
yield perturbation spectra consistent with
current CMB and LSS data
in the range of observable scales.

From these examples it can also be concluded that 
radically different inflationary dynamics
can yield the same asymptotic 
functional form for the tensor to scalar ratio. Therefore, the
derivation of the inflaton potential from the information on the
functional form of $r$ will hardly be unique. Here it is important to
note that by `different dynamics' it is implied solution behaviors,
which can be quite different for several values of the involved parameters,
even if the corresponding analytical expression for the potential, 
with unvalued parameters, is unique.

The above conclusion must be regarded together with the fact that 
it seems not possible to obtain information on the inflaton potential
beyond the exponential form,
using exclusively the observational information on values of
the spectral indices ($n_S$ and $n_T$)
and the running of the scalar index ($d n_S/ d\ln k$)
evaluated at the
pivot scale, or on a constant central value of the 
tensor to scalar ratio \cite{Terrero-Escalante:2002qe}. 
This would be a serious handicap for any program of reconstruction of the
inflaton potential.

A way to improve the 
situation described in the above paragraphs, could be to combine the 
information on 
$\Delta n$ (the difference between the spectral indices) and the value of $r$,
with the two first 
horizon--flow functions.
$\Delta n$ can be derived from observations and, according to
 Eq.~(\ref{eq:dlnr/dlnk}), gives information on the scale 
dependence of $r$. With this information evaluated at the pivot scale, 
the underlying inflaton potential
can be promptly sketched substituting  expression (\ref{eq:e1a1}) into 
Eq.~(\ref{eq:Vphi}). Moreover, here all three cases, power--law inflation,
$B\neq 0$ (if imitating power--law inflation) and $B=0$ can be differentiated. 
In the case of
power--law inflation, $\Delta n$ and $\epsilon_2$ will be
zero. If $B\neq 0$ then, as discussed in the previous section, 
its absolute value must be extraordinarily small what, 
after ten or more e--foldings, 
leads to $\epsilon_3\approx -1/C \approx 1.3712$, a
distinctively large, hardly suitable, value which will be the dominant 
contribution to
$\Delta n$ according to Eq.~(\ref{eq:dlnrdlnk1}).  
Finally, for $B=0$, $\epsilon_2=Cr\Delta n=A$, $\epsilon_3=0$ 
and the potential will be given by Eq.~(\ref{eq:Vai}).
As shown in Fig.~\ref{fig:01}, 
realizations of this potential
resemble the cases of monomial potentials with even order
($V_I$, $\epsilon_2<0$), 
and inflation near a maximum ($V_{II}$, $\epsilon_2>0$)
(see Ref.~\cite{inflation} for
examples of such inflationary scenarios)
allowing, therefore, to observe features of the inflaton potential beyond the
exponential form characteristic of power--law inflation. 

\begin{acknowledgments}
This research was supported in part by the CONACyT 
(Mexico) grant 38495--E. 
The author thanks A.\ Garc\'{\i}a, D.\ J.\ Schwarz,
A.\ de\ la\ Macorra and J.\ E.\ Lidsey
for strong support and helpful discussions.
\end{acknowledgments}


\end{document}